\documentclass[
twocolumn,prl,amsmath,amssymb]{revtex4}

\usepackage{graphicx}
\usepackage{amssymb}
\usepackage{amsmath}
\usepackage{dsfont}
\usepackage{bm}
\usepackage{mathrsfs}
\usepackage{times}

\begin{document}
\title{Derivation of quantum probabilities from deterministic evolution}

\author{T.\ G.\ Philbin}
\email{t.g.philbin@exeter.ac.uk}
\affiliation{Physics and Astronomy Department, University of Exeter,
Stocker Road, Exeter EX4 4QL, United Kingdom}

\begin{abstract}
The predictions of quantum mechanics are probabilistic. Quantum probabilities are extracted using a postulate of the theory called the Born rule, the status of which is central to the ``measurement problem" of quantum mechanics. Efforts to justify the Born rule from other physical principles, and thus elucidate the measurement process, have involved lengthy statistical or information-theoretic arguments. Here we show that Bohm's deterministic formulation of quantum mechanics allows the Born rule for measurements on a single system to be derived, without any statistical assumptions. We solve a simple example where the creation of an ensemble of identical quantum states, together with position measurements on those states, are described by Bohm's quantum dynamics. The calculated measurement outcomes agree with the Born-rule probabilities, which are thus a consequence of deterministic evolution. Our results demonstrate that quantum probabilities can emerge from simple dynamical laws alone, and they support the view that there is no underlying indeterminism in quantum phenomena.
\end{abstract}

\maketitle

\section{Introduction}
In the usual textbook presentation, quantum mechanics is an indeterministic theory whose predictions are statistical~\cite{merzbacher}. This differs fundamentally from classical mechanics, in which the dynamical laws are deterministic but can sometimes lead to statistical predictions~\cite{LL}. In quantum mechanics the Schr\"{o}dinger equation describes deterministic (unitary) time evolution. The indeterministic and probabilistic aspect of quantum mechanics enters with the Born rule; in its original form~\cite{bor26} the rule associates the square of the absolute value of the wave function with a probability density for position, provided a position measurement is performed. The Born rule is an additional postulate essential for obtaining experimental predictions, but its requirement to specify which events constitute measurements has long been a source of dissatisfaction and controversy~\cite{bell}. Efforts to base the theory solely on the dynamics of the Schr\"{o}dinger equation face the well-known crux of reconciling strict unitary evolution with the classical-like behaviour of macroscopic objects~\cite{bell,deu99,zur09}. There is no general agreement  that the emergence of classical behaviour from the quantum world could eventually be explained by unitary evolution~\cite{sch13}.

The status of the Born rule requires reassessment in the formulation of non-relativistic quantum mechanics developed by de Broglie, Madelung and Bohm, generally referred to as Bohmian mechanics~\cite{deb28,mad27,boh52,bohm,holland,duerr}. In this formulation, quantum mechanics is a deterministic theory in which particles follow trajectories just as in classical mechanics. The Schr\"{o}dinger equation is supplemented by a simple equation relating the particle trajectory to the phase of the wave function (equation~(\ref{x}) below). The statistical aspect of quantum mechanics is incorporated by postulating that the particle trajectories in an ensemble of identical quantum states are distributed with a probability density given by the square of the absolute value of the wave function. Once such a distribution is established it is preserved for all later times by the Bohmian dynamical laws~\cite{boh52,bohm,holland,duerr}. This formulation gives quantum mechanics more of the features of classical statistical mechanics, and it naturally leads to the  question of why the particle trajectories should be distributed in any particular way. One point of view that arises from the similarities to statistical mechanics is that there exists a natural or typical distribution of the trajectories, just as thermal-equilibrium distributions are viewed as typical in statistical mechanics~\cite{duerr,val91,due92}. This requires statistical arguments and measures of typicality for large systems subject to the (deterministic) Bohmian dynamical laws~\cite{duerr,val91,due92}. Here we first show that consistency of the Bohmian formulation requires the Born rule for measurements on a single system to be derivable from the deterministic evolution, without any appeal to statistical arguments. Then we give such a derivation of the Born rule for a particular system. 

\section{General argument} \label{sec:gen}
The experimental verification of the Born rule for measurement outcomes requires an ensemble of identical quantum states. Each state in the ensemble can be prepared using different physical systems with measurements performed separately on each system. Alternatively, the same physical system can be employed throughout, with the system being returned to the quantum state in question after each measurement. We focus on the second scenario from now on. The significance of this second scenario is that the entire process of measurements and re-formations of the quantum state can be described using the deterministic laws of Bohmian quantum mechanics. Reproduction of the original quantum state after a measurement is described by the Schr\"{o}dinger equation with an appropriate external potential, and this process is deterministic even in the usual formulation of quantum mechanics. But in the Bohmian formulation the measurement process itself is also deterministic, with the outcome of a measurement determined by the actual position of the particle in the quantum state~\cite{boh52,bohm,holland,duerr}. The statistical distribution of experimental results is then usually attributed to the postulated distribution of the particle trajectories in an ensemble of identical states~\cite{boh52,bohm,holland,duerr}. In our scenario, however, a complete description of the ensemble of identical recurring states and measurements on those states will give us the particle trajectory throughout the entire process. We are thus not free to assume any statistical distribution for the position of the particle in the recurring states: our calculation will give us all of these positions. The position in each recurrence of the quantum state will in turn give us the result of the measurement on that state, so we will have derived the measurement outcomes for the ensemble from deterministic laws. If these measurement results exhibit the Born rule then we will have derived this rule from the Bohmian formulation; if the results do not obey the Born rule then this rule is inconsistent with Bohmian quantum mechanics. The only degree of freedom of the particle trajectory in the ensemble of recurrences is the particle position in the initial state~\cite{boh52,bohm,holland,duerr}. In the Bohmian formulation the particle can be at any position where the wave function is non-zero (but in any particular state the particle is in a definite position)~\cite{boh52,bohm,holland,duerr}. If we are to derive the Born rule, the derivation must hold for any allowed position of the particle in the initial state because the Born rule must hold for any ensemble of recurrences and measurements.

These considerations show that the Born rule for measurement outcomes must be derivable in Bohmian quantum mechanics without any statistical assumptions; it cannot be an extra postulate nor can it be traced to some typicality property of a universe obeying quantum mechanics, even if the latter notion could be made meaningful given the unsolved problem of quantum gravity. 

This general argument has already been given in~\cite{sht97,gei02}, but without an explicit demonstration that the Born rule can indeed be derived in the manner described. Instead appeals were made to a likely ergodicity~\cite{sht97} or chaotic dynamics~\cite{gei02} in the system trajectory when repeated measurements are performed. However, what is required for a demonstration is an explicit  \emph{calculation of measurement outcomes} using only the deterministic dynamical equations, and this is what will be provided below. It should be noted that the difficult aspect of any such calculation is the requirement that the (deterministic) evolution of the system and measuring device must feature repeated measurements of the system. Solutions of the Bohmian equations that show recurrences of the same quantum state will in general have no relevance to the Born rule for measurement outcomes and will not exhibit any behaviour that could be called "ergodic". Only when the recurring quantum state is subjected to repeated measurements (and these measurements are fully included in the dynamics) can the Born rule be tested. In our example the repeated measurements will be rather easily included because of the details of the evolution but in general the calculation of the measurement outcomes will be very difficult.

\begin{figure}[h!]
\centering
\includegraphics[width=8.6cm]{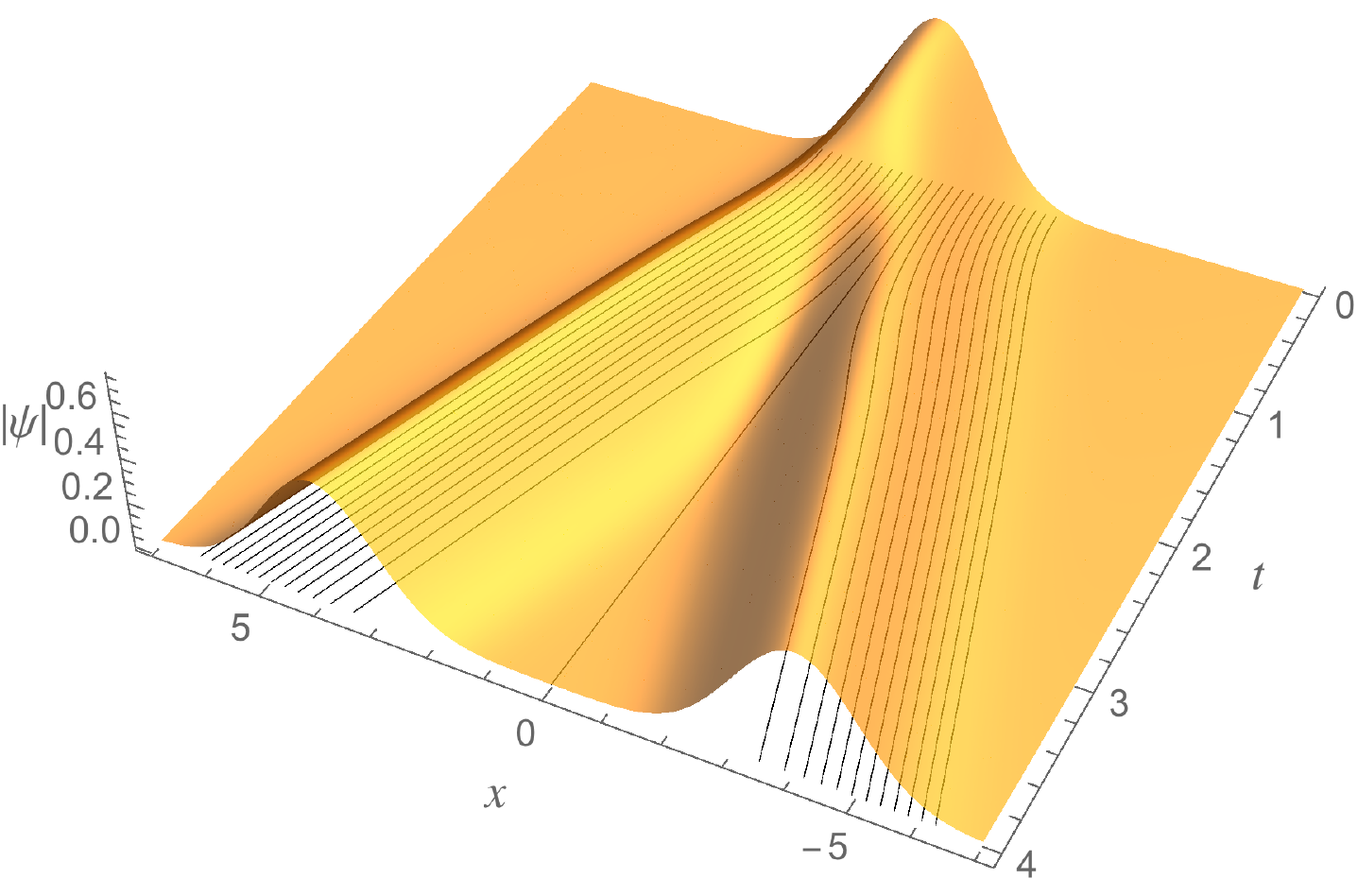}
\caption{The amplitude (\ref{R}) of the wave function (\ref{psi}). The black lines in the $xt$-plane are trajectories of the particle for a range of initial positions (see (\ref{x})).}
\label{fig:psi}
\end{figure}

\section{Example}
\begin{figure}[h!]
\centering
\includegraphics[width=8.6cm]{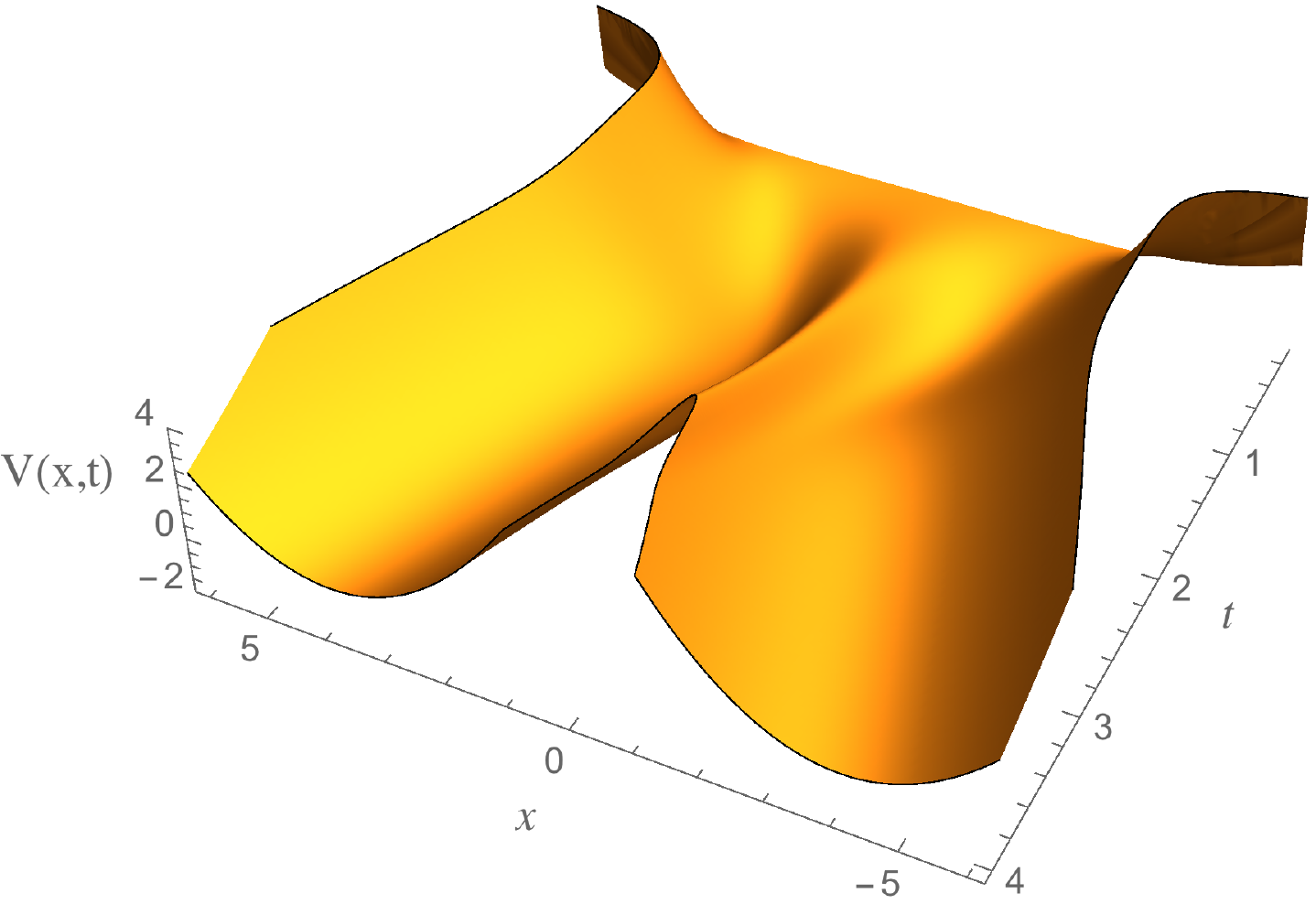}
\caption{The external potential (\ref{V}), which vanishes at $t=0$ and forms two harmonic-oscillator potentials moving in opposite directions at unit speed as $t\to\infty$.}
\label{fig:V}
\end{figure}

The preceding argument can be illustrated and tested by an example. We choose the state for the ensemble to be the ground state of the one-dimensional harmonic oscillator. The measurement performed on the state will be a determination of whether the particle lies to the left or the right of the peak of $|\psi|^2$ (a Gaussian function of position). This measurement process requires us to split the wave function into two equal parts (similar to a Stern-Gerlach experiment~\cite{merzbacher}). We must describe this splitting dynamically using the (deterministic) laws of Bohmian quantum mechanics. First we identify an external potential that performs the required splitting of the wave function. A simple approach is to write down a time-dependent wave function $\psi(x,t)$ that exhibits the splitting in question, substitute this into the Schr\"{o}dinger equation, and solve for the potential. Consider the wave function
\begin{gather}
\psi(x,t)=R(x,t)e^{iS(x,t)},  \label{psi}  \\
R(x,t)=\left(\frac{4}{\pi}\right)^{\frac{1}{4}}  \frac{e^{-x^2/2}\cosh (xt)}{\sqrt{e^{t^2}+1}},  \label{R}  \\
S(x,t)=f(x,t)-\frac{t}{2},  \label{S} 
\end{gather}
where $R(x,t)$ and $S(x,t)$ are the (real) amplitude and phase, respectively. Throughout this paper we take $\hbar=1$ and consider a particle of unit mass. The amplitude (\ref{R}) is plotted in Fig.~\ref{fig:psi}. It is a Gaussian at $t=0$ that splits into two Gaussians moving in opposite directions with unit speed as $t\to\infty$. We require the external potential $V(x,t)$ that produces this dynamics to be real and this will determine the unspecified real function $f(x,t)$ in the phase (\ref{S}), as follows. Substitution of (\ref{psi})--(\ref{S}) into the Schr\"{o}dinger equation yields a complex expression for $V(x,t)$; setting the imaginary part of this expression to zero gives a differential equation for $f(x,t)$. The differential equation features only spatial derivatives of $f(x,t)$, up to second order, and a solution is
\begin{align} 
 &f(x, t)  =  \int_0^x dy {\Bigg [} \tanh (yt)   \nonumber \\
 & + \left. \frac{\sqrt{\pi} \, t \, e^{(y+t)^2}[\mathrm{erf}(t-y)+2\mathrm{erf}(y)-\mathrm{erf}(t+y)]}{(e^{t^2}+1)(e^{2yt}+1)^2} \right],     \label{f}   
\end{align}
where $\mathrm{erf}(x)=(2/\sqrt{\pi}) \int_0^x dy\, e^{-y^2}$ is the error function. The (real) external potential is then given by
\begin{equation} \label{V}
V(x,t)=\frac{x^2+t^2}{2}-xt\tanh(xt)-\partial_tf(x,t)-\frac{1}{2}[\partial_xf(x,t)]^2.
\end{equation}
and this is plotted in Fig.~\ref{fig:V}. The potential is zero at $t=0$ and as $t\to\infty$ it becomes two harmonic oscillator potentials (for unit oscillation frequency) moving in opposite directions with unit speed. The initial ($t=0$) vanishing of the potential $V(x,t)$ causes the wave function to begin to spread; the subsequent ($t>0$) behaviour of $V(x,t)$ marshals the wave function in the correct manner to produce the splitting shown in Fig.~\ref{fig:psi}. 

\begin{figure}[h!]
\centering
\includegraphics[width=8.6cm]{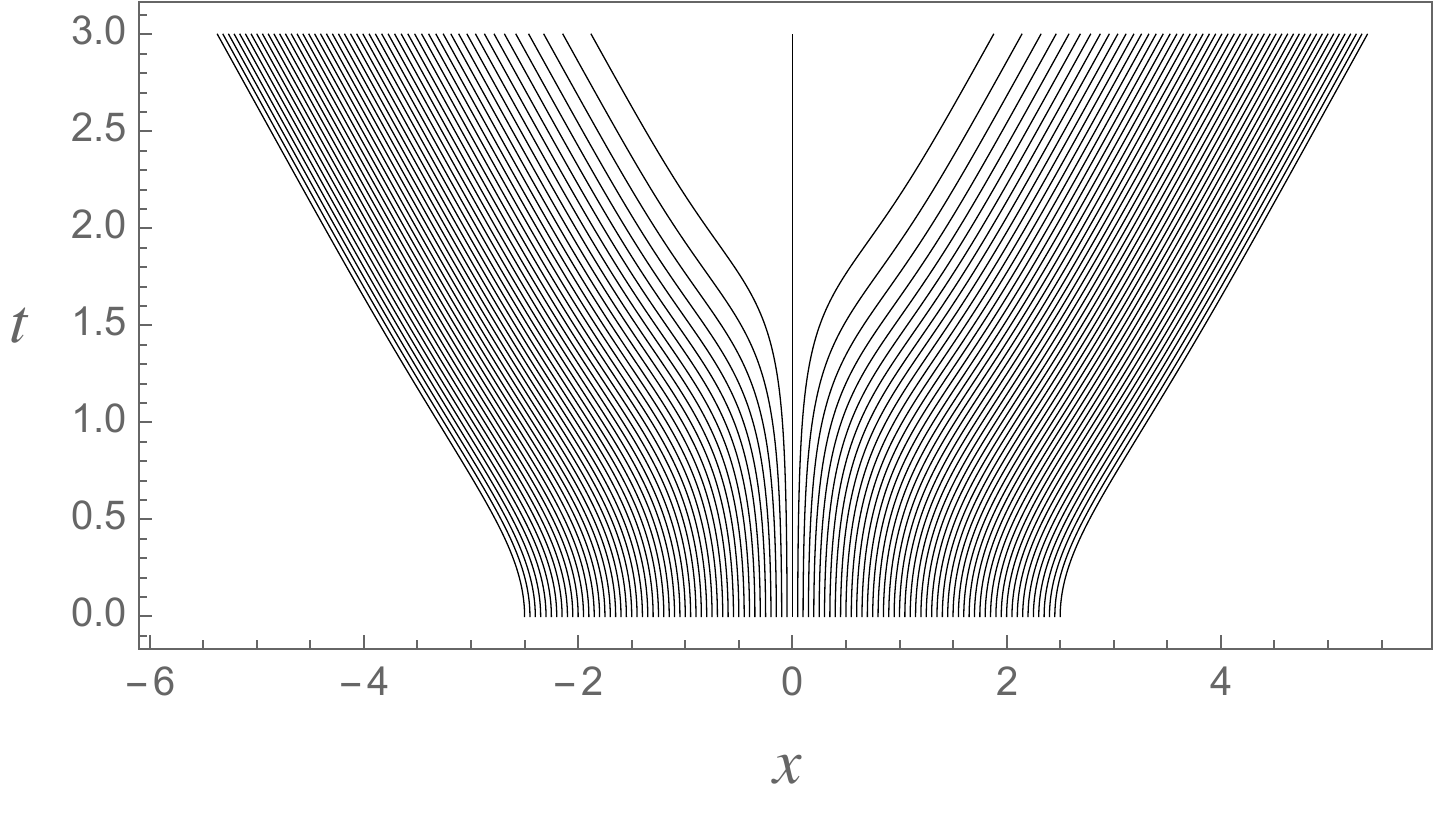}
\caption{Possible particle trajectories in the state (\ref{psi})--(\ref{S}). The trajectories are solutions of (\ref{x}) for a range of initial positions $x(0)$.}
\label{fig:trajs}
\end{figure}

The particle in the state (\ref{psi})--(\ref{S}) follows one of an infinite number of possible trajectories determined by its position at any time, say $t=0$. The trajectories are the solutions of
\begin{equation}  \label{x}
\dot{x}(t)=\left. \partial_xS(x,t)\right|_{x=x(t)},
\end{equation}
with an arbitrary initial position $x(0)$~\cite{boh52,bohm,holland,duerr}. Sample trajectories are plotted in Figs.~\ref{fig:psi} and~\ref{fig:trajs} for a set of initial positions. Depending on whether the particle has initial position $x(0)<0$ or $x(0)>0$ it moves to the left or right, respectively, in Fig.~\ref{fig:trajs}. The particle initially has zero velocity and is accelerated until it is moving left or right with unit speed (see Fig.~\ref{fig:trajs}). For initial position $x(0)=0$  the particle remains at rest; this is a point of unstable equilibrium. 

It is essential for our argument that no assumptions are made about which trajectory the particle follows. In particular we must not attribute any statistical meaning to the wave function. Although we have followed convention in using a normalized wave function (\ref{psi})--(\ref{S}) (i.e.\ $\int_\infty^\infty dx\,|\psi|^2=1$), this is irrelevant for what follows: the trajectories are independent of any normalization factor (as is obvious from (\ref{x})).

To complete the position measurement on the initial state we must ascertain whether the particle is moving left or right, as this reveals whether the particle in the initial state was located at $x<0$ or $x>0$, respectively (this statement holds in both the usual and Bohmian formulations). This requires the coupling of the particle to a macroscopic ``pointer"~\cite{bell,boh52,bohm,holland,duerr}, which in this case will take one of two values (left/right). In the usual formulation of quantum mechanics the pointer is a classical object and when the position of the pointer is read the quantum state of the particle ``collapses" irreversibly to the corresponding part (left or right) of the split wave function~\cite{bell}. In Bohmian mechanics both the pointer and the particle obey the quantum dynamical laws and there is no collapse; rather the measurement process entangles the pointer with the measured system so that movement of the pointer to ``left" or ``right" shows that the particle is in a left or right trajectory, respectively. This simple account of measurement is possible in the Bohmian formulation because quantum objects (pointer and particle) follow well-defined, unique trajectories in space~\cite{boh52,bell,bohm,holland,duerr}. Once it is ascertained that the particle is moving left, say, the right-moving part of the split wave function no longer affects the particle trajectory, but one might envisage bringing the "empty" right-moving part of the wave function back into contact with the partner from which it split at some point in the future. The pointer's wave function has undergone a similar splitting however, and its coupling to the wider environment produces a total system wave function (pointer, particle and environment) that is a superposition of non-overlapping functions in the total configuration space~\cite{boh52,bohm,holland,duerr}. The total system follows a definite trajectory in its configuration space, located in one of the non-overlapping parts of its wave function~\cite{boh52,bohm,holland,duerr}. Subsequent overlapping of the constituent parts of the superposition is possible in principle but viewed as overwhelmingly unlikely in practice and this is how the irreversibility and time asymmetry of the measurement process enters into the Bohmian formulation~\cite{boh52,bohm,holland,duerr}. In our example this means that the two split parts of the particle wave function cannot subsequently interfere (even if brought back together) because the corresponding parts of the pointer-plus-environment wave function do not overlap.  Thus, after the splitting \emph{and} the measurement, the empty part of the particle wave function can have no further effect on the particle motion.

After the coupling to a left/right pointer we must re-form the original quantum state from the part (left- or right-moving) of the wave function relevant for the subsequent particle motion. Examination of (\ref{psi})--(\ref{f}) shows that as $t\to\infty$ each of the split parts of the wave function reproduces the original $\psi(x,0)$ in a frame moving at constant unit speed with amplitude reduced by a factor of $1/\sqrt{2}$ and with an additional time-dependent phase factor. In more detail, the phase function (\ref{f}) is zero at $t=0$ and approaches $x$ for $x>0$, $-x$ for $x<0$, as $t\to\infty$; this gives the particle a unit speed as $t\to\infty$, from (\ref{x}). The resulting left/right wave packets are Galilean transformations~\cite{merzbacher} of the ground state wave function of the harmonic oscillator up to a time-dependent phase factor (and a factor of $1/\sqrt{2}$).  Any $x$-independent terms in the phase have no effect on the particle motion (see (\ref{x})) and neither do constant factors in the wave function. Thus the left- and right-moving wave packets both reproduce the original $t=0$ quantum state in a different inertial frame. After the coupling to the left/right pointer we must therefore undo any change in the quantum state caused by this coupling in order to reproduce the original state. For simplicity we assume that an energy measurement is performed at large time $t$ in each of the two harmonic-oscillator potentials produced by the evolving external potential (\ref{V}): one of these measurements will yield the ground-state energy and leave the quantum state undisturbed while the other will find no particle in the well (the pointer will not move). The result of the left/right measurement is the direction (left or right) taken by the particle, which in turn is completely determined by the initial position $x(0)$. After the measurement the empty part of the wave function is irrelevant (see above) and the particle is now known to be located in a wave function (left- or right-moving) that reproduces the original wave function (up to irrelevant constant factors and time-dependent phase factors). It should be noted again that the measurement just described is an essential requirement in our calculation because without measurement outcomes we cannot derive the Born rule. The relative ease with which measurement has been incorporated is due to the details of this particular example; in general the inclusion of measurements in the evolution will be  a difficult task.

\begin{figure}[h!]
\centering
\includegraphics[width=8.6cm]{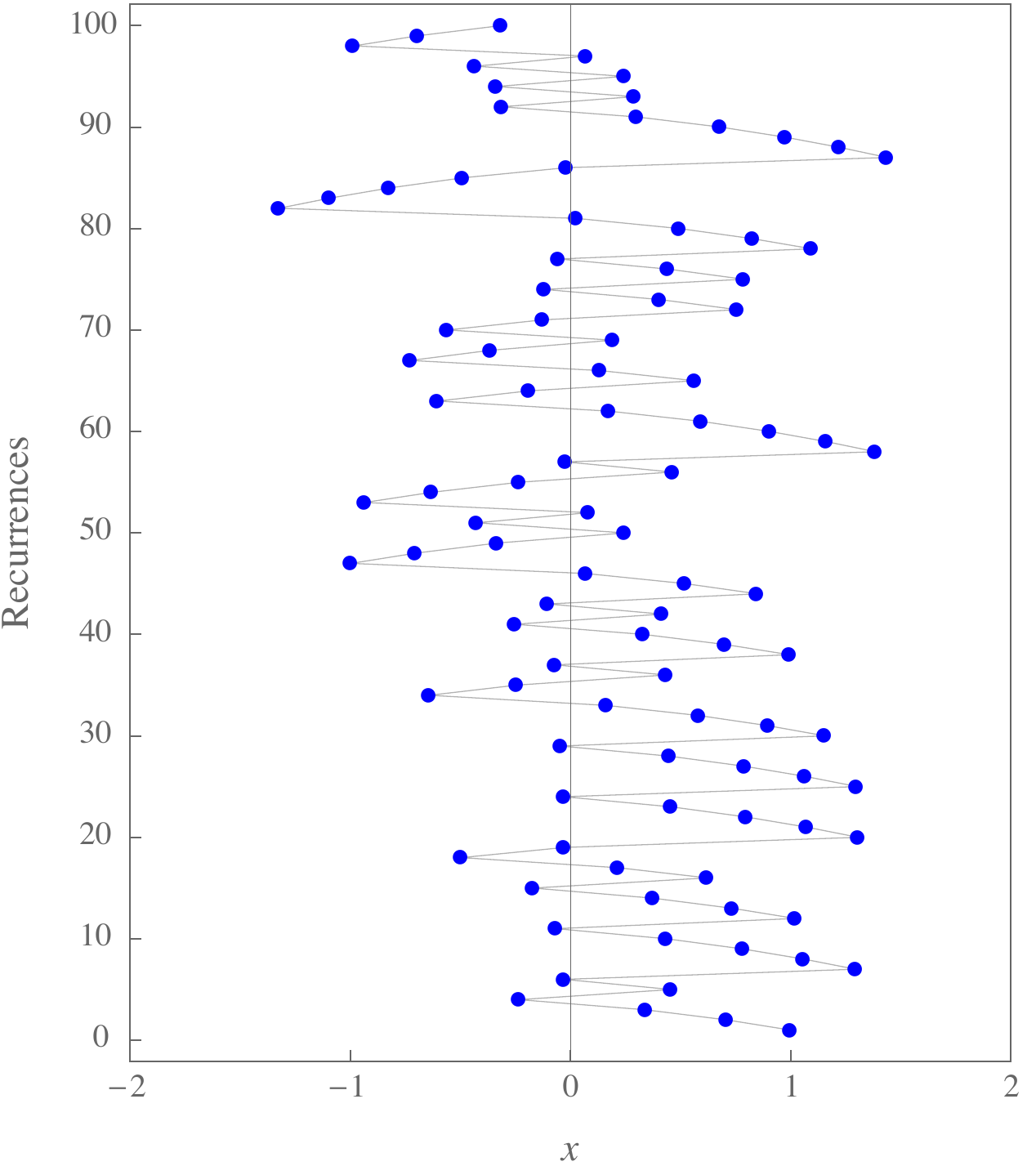}
\caption{Position of the particle relative to the centre of the harmonic-oscillator well for the first 100 recurrences of the quantum state. The initial position is $1$ and the trajectory equation (\ref{x}) is solved from $t=0$ to $t=20$ between recurrences. Consecutive positions are connected by a straight grey line.}
\label{fig:pts}
\end{figure}

The post-measurement recurrence of the original quantum state gives the second member of the ensemble. By transferring to the inertial frame co-moving with the particle we can repeat the process just described; in this manner we generate the ensemble of identical quantum states and measurement results on those states. For each value of the initial position $x(0)$, the particle trajectory will give the measurement outcomes on every state in the ensemble without any use of the Born rule or statistical arguments. We can thus check if the measurement outcomes agree with the Born rule for every $x(0)$. The relevant property of the trajectory is the particle position in each recurrence relative to the centre of the harmonic-oscillator well in which it lies, as this gives  the measurement outcomes. We can employ a simple iterative procedure: we choose $x(0)$, solve the trajectory equation (\ref{x}) (see Fig.~\ref{fig:trajs} for examples) to find $x(t)$ at large $t$, subtract from $x(t)$ the position of the centre of the harmonic-oscillator well in which the particle lies, and use the resulting number as the initial position in the next iteration. A sample set of relative positions obtained from such an iteration is shown in Fig.~\ref{fig:pts} for the choice $x(0)=1$ and with the trajectory equation solved from $t=0$ to $t=20$ to obtain the $(n+1)$st position from the $n$th position. The qualitative behaviour of consecutive positions in Fig.~\ref{fig:pts} can be understood from the trajectories in Fig.~\ref{fig:trajs}. A position far from the centre of the well is moved back towards the centre, but a position very close to the centre is followed by a position far from the centre on the opposite side. This entirely deterministic process produces a distribution of positions whose statistics we now investigate.

In the usual formulation of quantum mechanics, the measurements on the ensemble give a sequence of ``left" or ``right" readings and the Born rule predicts an equal likelihood for each of the two possible readings. Our sequence of relative positions $\{x_i\}$ in the recurrences gives these measurement results according to the correspondence $x_i<0\Rightarrow\mathrm{LEFT}$, $x_i>0\Rightarrow\mathrm{RIGHT}$. We can check whether the statistics of $\{x_i\}$ agrees with this Born-rule prediction. One might imagine that a more detailed test of the Born rule (involving more than two possible measurement outcomes) would be obtained by splitting the wave function into a very large number of parts and then coupling each part to a pointer to perform a much more exact position measurement. In fact the left/right measurement data already allows us to determine the particle position in the initial state, and indeed in any of the recurrences, to arbitrary precision. This is because our derivation of the measurement outcomes is only possible using the Bohmian formulation, and the sequence of left/right measurement results allows us to locate more and more precisely which trajectory the particle is following. The result of the $n$th left/right measurement is determined by the initial position $x(0)$ and for large $n$ the range of $x(0)$ that corresponds to trajectories that reach the $n$th recurrence is very small (only one sequence of left/right deflections of the trajectory leads to this recurrence). The initial position $x(0)$ can thus be determined to arbitrary accuracy (on this kind of retrodiction in Bohmian mechanics, see Ref.~\onlinecite{holland}). Moreover the same is true for the particle position in any recurrence: by extending the evolution to additional recurrences we can determine the position in any particular recurrence to arbitrary precision. Hence our set $\{x_i\}$ can be taken to be position measurements on an ensemble of identical quantum states and the statistics of these measurements has thus been derived from the deterministic laws of Bohmian mechanics. 

\begin{figure}[h!]
\centering
\includegraphics[width=8.6cm]{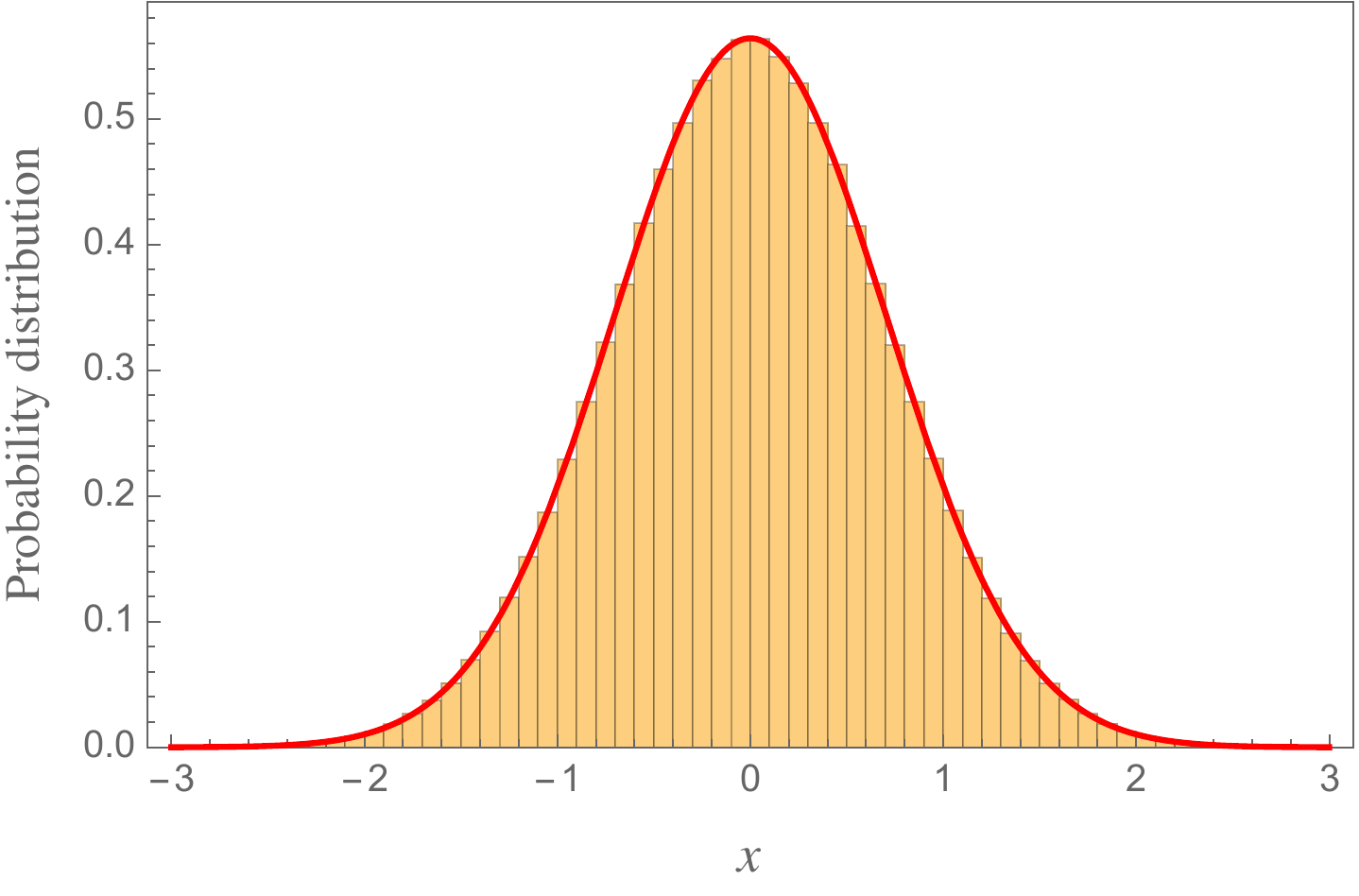}
\caption{Statistics of particle positions in the ensemble. $2\times 10^6$ positions are used, the first 100 of which are shown in Fig.~\ref{fig:pts}. The histogram shows the number of points in bins of width 0.1, normalized to give the probability density function. The red curve is the (normalized) $|\psi|^2$ for the ensemble.}
\label{fig:stats}
\end{figure}

The statistics of the particle positions in the ensemble is displayed in Fig.~\ref{fig:stats}, for the choice of $x(0)$ and evolution time between recurrences used in Fig.~\ref{fig:pts}. The particle positions in two million recurrences were calculated and grouped into bins of width 0.1. Figure.~\ref{fig:stats} shows a histogram of the number of points in each bin, normalized to give the probability density function. The red curve in Fig.~\ref{fig:stats} is the (normalized) $|\psi|^2$ for the quantum state in the ensemble. Different choices of initial position $x(0)$ do not affect the statistics (the choice $x(0)=0$ is an exception, as noted earlier, but infinitesimal differences in the quantum states in the recurrences would move the particle off this unstable trajectory even if the initial position were exactly $x(0)=0$). In particular a choice of $x(0)$ at which the value of $|\psi|^2$ is negligible produces the same histogram for large number of recurrences. As noted above, this lack of dependence of the distribution on the initial position is essential if the Born rule is to be a consequence of Bohmian quantum mechanics. Another requirement is that minor differences in the quantum states of members of the ensemble should lead to only minor changes in the distribution. This requirement is also verified to some extent by our example because the evolving wave function (\ref{psi})--(\ref{S}) only reproduces the initial $t=0$ quantum state asymptotically as $t\to\infty$. By halting the evolution after finite time and taking the left or right part to be a recurrence of the original state, as is done in Figs.~\ref{fig:pts} and~\ref{fig:stats}, we introduce small errors. Such tiny differences in the quantum states are in fact important for the position of the particle in individual recurrences: a change in the time for which the trajectory equation is integrated significantly changes the particle positions after a few tens of recurrences. The distribution of positions is unaffected however, and this stability is essential for the distribution to have experimental significance.

\section{Conclusion}
We have given an example where the Born rule for quantum probabilities is a consequence of deterministic evolution, with no statistical assumptions. A second, closely related example is given in the Appendix. These examples show ergodic features in the Bohmian trajectories and it was argued in~\cite{sht97,gei02} that this should be the case for an ensemble of recurrences and measurements. 

These results naturally lead to the hypothesis that all ensembles of recurrences and measurements in the Bohmian formulation will exhibit the Born rule for measurement outcomes. Indeed this must be so if Bohmian quantum mechanics is to agree with experiment. We stress again that the evolution must incorporate measurements on the recurring quantum state; if measurements are not included in the dynamics then there is nothing to compare with the Born rule (which refers to measurement outcomes). If this hypothesis is correct, then Bohm's approach provides the simplest resolution of the measurement problem in the case of measurements on a single system. Bohm's theory, strengthened by our findings, supports the view that indeterminism in the quantum world is a theoretical choice~\cite{bell}, not an experimental fact.

\section*{Acknowledgements}
I thank S.~Horsley and J.~Anders for helpful suggestions.

\vspace{5mm}

\appendix*
\section{Appendix}

\begin{figure}[h!]
\centering
\includegraphics[width=8.6cm]{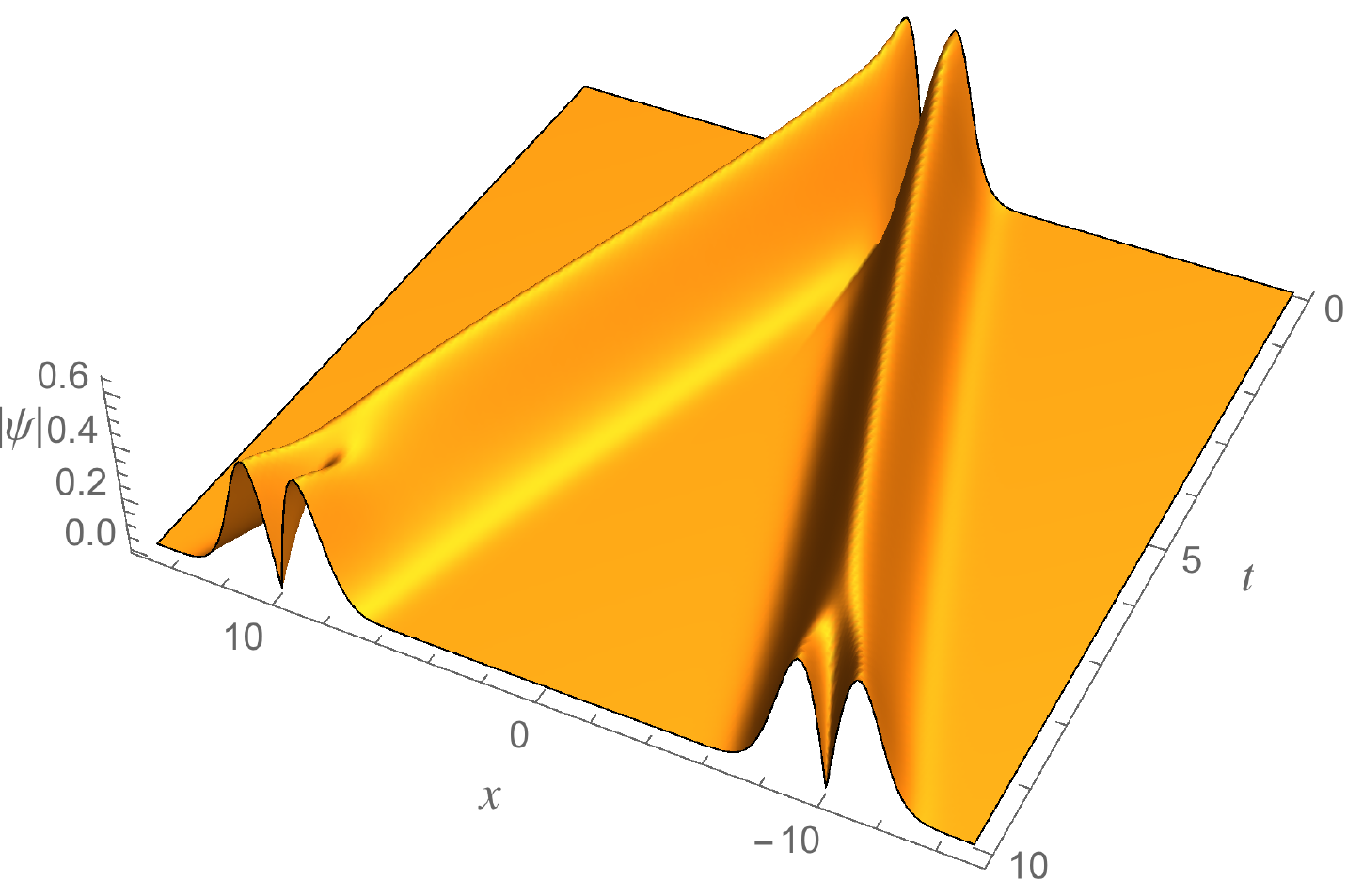}
\caption{ A splitting of the wave function for the first excited state of the harmonic oscillator. The amplitude $R(x,t)=|\psi(x,t)|$ of the wave function (\ref{psi}) is shown.}
\label{fig:psia}
\end{figure}

Here we show a second example of an ensemble of recurrences and measurements. This second example is similar to that in the main text, but here we use the first excited state of the harmonic oscillator as the recurring state in the ensemble. We chose a normalized wave function (as in the main text, $\hbar=m=1$)
\begin{widetext}
\begin{equation} \label{psia}
\psi(x,t)=\frac{e^{if(x,t)-3 i t/2}\left[e^{-(x-t)^2/2} (t (x-t)+i (10-t) x)+e^{-(t+x)^2/2} (t (t+x)+i (10-t) x)\right]}{(4\pi)^{1/4}e^{-t^2/2} \sqrt{t \left(-t^3+t-10\right)+e^{t^2} [(t-10) t (t^2-10 t+1)+50]+50}},
\end{equation}
\end{widetext}
where $f(x,t)$ is an unknown (real) phase. The amplitude $R(x,t)=|\psi(x,t)|$ of this wave function is shown in Fig.~\ref{fig:psia}.

\begin{figure}[h!]
\centering
\includegraphics[width=8.6cm]{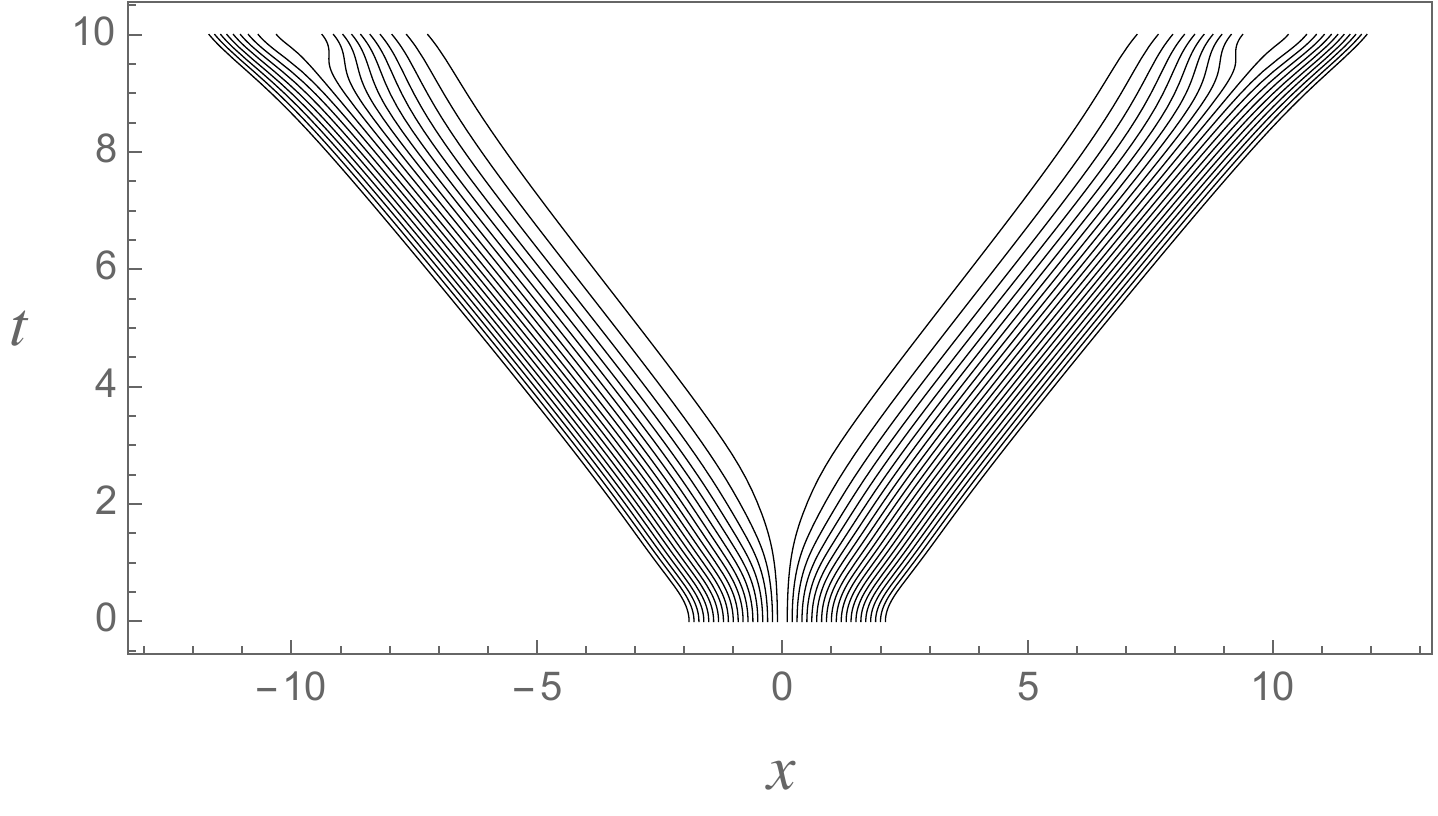}
\caption{ Possible particle trajectories in the splitting of the first excited state of the harmonic oscillator.}
\label{fig:trajsa}
\end{figure}

The amplitude of the wave function (\ref{psia}) at $t=0$ is that of the first excited state of a harmonic oscillator. At $t=10$ the amplitude is in two spatially separated parts, each of which is, to a very close approximation, also the amplitude of the first excited state (up to a constant factor). Unlike the example in the main text, here the re-formation of the original state after the splitting is not arranged to be exact in the asymptotic limit $t\to\infty$, rather we arrange an almost exact  re-formation at a finite time (here $t=10$). For $t>10$ the wave function (\ref{psia}) does not reproduce the initial amplitude. The phase $f(x,t)$ in (\ref{psia}) must be chosen so that the external potential $V(x,t)$ is real. Substituting (\ref{psia}) into the Schr\"{o}dinger equation and solving for $V(x,t)$ we demand that the resulting expression be real. This determines $f(x,t)$ up to a additive constant. In detail, $f(x,t)$ must be related to the amplitude $R(x,t)$ of the wave function  by~\footnote{I am indebted to S.~Horsley for pointing out this simple relation.}
\begin{equation}  \label{frel}
\partial_x f(x,t)=-\frac{1}{R^2(x,t)}\int_0^x dx\, \partial_t\left[R^2(x,t)\right].
\end{equation}
Using (\ref{frel}) an analytic expression for $\partial_x f(x,t)$ can be found using Mathematica, but it is very lengthy and we do not write it here. Since the spatial derivative of the phase determines the particle trajectories (see main text) we obtain an exact analytical equation for the trajectories in this second example. Samples of trajectories are plotted in Fig.~\ref{fig:trajsa} for a set of initial positions $x(0)$.

\begin{figure}[h!]
\centering
\includegraphics[width=8.6cm]{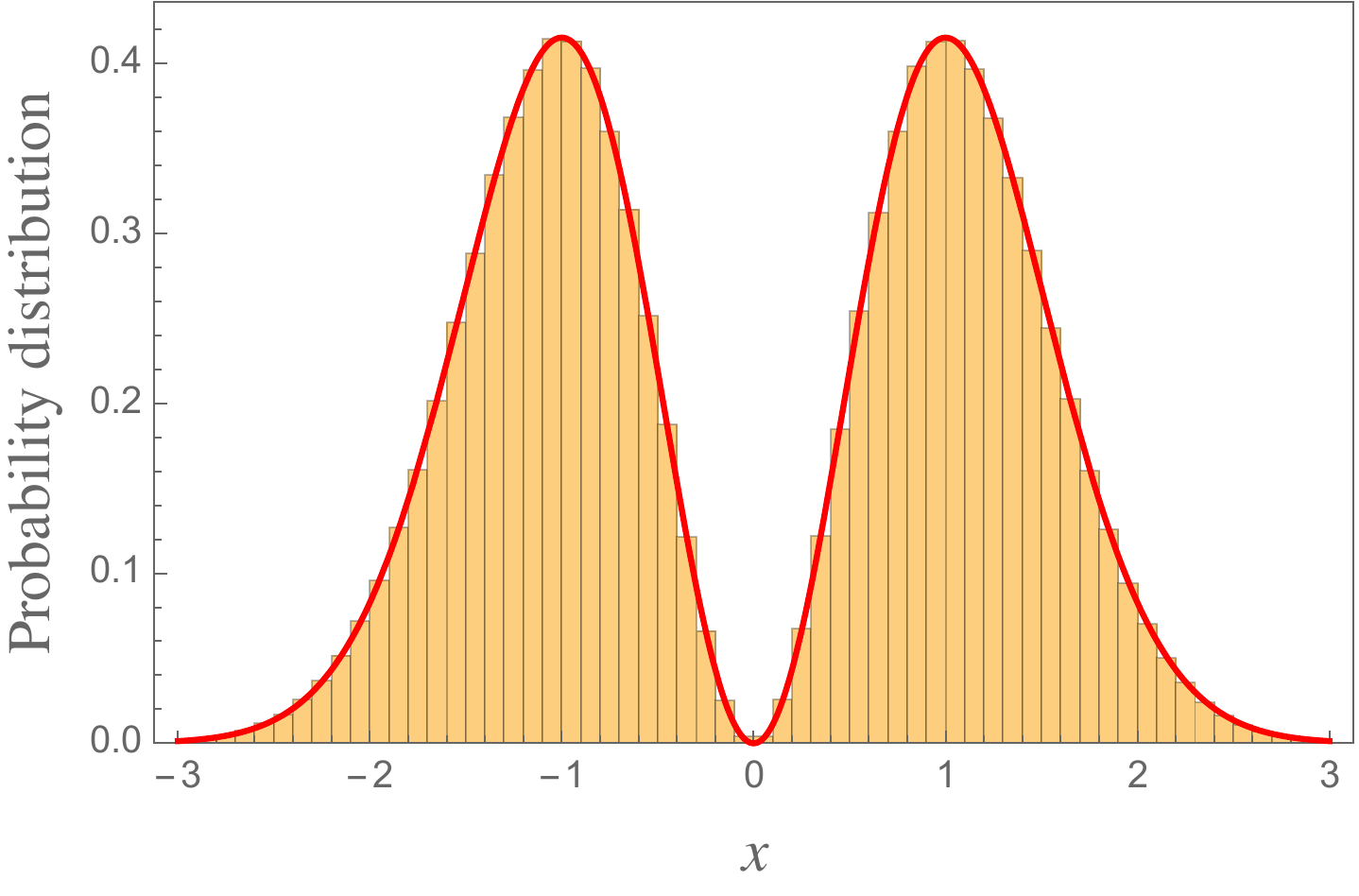}
\caption{ Statistics of particle positions in the ensemble of first excited states of the harmonic oscillator. $10^5$ positions were calculated for initial position $x(0)=1$. The histogram shows the number of points in bins of width 0.1, normalized to give the probability density function. The red curve is the (normalized) $|\psi|^2$ for the ensemble.}
\label{fig:dataa}
\end{figure}

We now perform the same procedure described in the main text. Here we integrate the trajectory equation from $t=0$ to $t=10$ between recurrences. This gives us the position of the particle relative to the centre of the wave function in each recurrence after a splitting. As described in the main text, these relative positions determine the results of the position measurements on the ensemble of recurrences. The statistics of the relative positions is depicted in Fig.~\ref{fig:dataa} for $10^5$ recurrences, with initial position $x(0)=1$. The positions were grouped into bins of width $0.1$, and Fig.~\ref{fig:dataa} shows a histogram of the number of points in each bin, normalized to give the probability density function. Also shown (red curve) in Fig.~\ref{fig:dataa} is the normalized $|\psi|^2$ for the state in the ensemble, which correctly gives the probabilities for the calculated measurement outcomes.


\begin{thebibliography}{99}

\bibitem{merzbacher}
Merzbacher, E. {\it Quantum Mechanics} 3rd ed.\ (Wiley, New York, 1998).

\bibitem{LL}
Landau, L.D.\ \& Lifshitz, E.M. \textit{Statistical Physics Part I} 3rd ed.\ (Butterworth-Heinemann, Oxford, 1980).

\bibitem{bor26}
Born, M. Zur Quantenmechanik der Sto{\ss}vorg\"{a}nge. \textit{Z.\ Phys.}\ \textbf{37,} 863 (1926). 

\bibitem{bell}
Bell, J.S. \textit{Speakable and Unspeakable in Quantum Mechanics} (Cambridge Univ.\ Press, 1987).

\bibitem{deu99}
Deutsch, D. Quantum theory of probability and decisions. \textit{Proc.\ R.\ Soc.\ A} \textbf{455,} 3129 (1999). 

\bibitem{zur09}
Zurek, W.H. Quantum Darwinism. \textit{Nature Phys.}\ \textbf{5,} 181 (2009). 

\bibitem{sch13}
Schlosshauer, M., Kofler, J.\ \& Zeilinger, A. A snapshot of foundational attitudes toward quantum mechanics. \textit{Stud.\ Hist.\ Phil.\ Mod.\ Phys.}\ \textbf{44,} 222 (2013). 

\bibitem{deb28}
de Broglie, L. La nouvelle dynamique des quanta. In \textit{\'{E}Žlectrons et photons: rapports et discussions du cinqui\`{e}me conseil de physique} (Gautier-Villars, Paris, 1928).

\bibitem{mad27}
Madelung, E. Quantentheorie in hydrodynamischer Form. \textit{Z.\ Phys.}\ {\bf 40,} 322 (1927).

\bibitem{boh52}
Bohm, D. A suggested interpretation of the quantum theory in terms of ``hidden" variables. \textit{Phys.\ Rev.}\ {\bf 85,} 166 (1952); \ {\bf 85,} 180 (1952).

\bibitem{bohm}
Bohm, D.\ \& Hiley, B.J. \textit{The Undivided Universe} (Routledge, London, 1993).

\bibitem{holland}
Holland, P.R. \textit{The Quantum Theory of Motion} (Cambridge Univ. Press, 1993).

\bibitem{duerr}
D\"{u}rr D.\ \& Teufel, S. \textit{Bohmian Mechanics} (Springer, Berlin, 2009).

\bibitem{val91}
Valentini, A. Signal-locality, uncertainty, and the subquantum \textit{H}-theorem. \textit{Phys.\ Lett.\ A} {\bf 156,} 5 (1991).

\bibitem{due92}
D\"{u}rr, D., Goldstein, S.\ \& Zangh\'{i}, N. Quantum equilibrium and the origin of absolute uncertainty. \textit{J.\ Stat.\ Phys.}\ {\bf 67,} 843 (1992).

\bibitem{sht97}
Shtanov, Yu.V. Origin of quantum randomness in the pilot wave quantum mechanics. arXiv:quant-ph/9705024.

\bibitem{gei02}
Geiger, H., Obermair, G.\ \& Helm, C. Quantum mechanics without statistical postulates. In \textit{Quantum Communication, Computing, and Measurement 3} (Springer, Berlin, 2002).

\end{thebibliography}
\end{document}